\documentclass[conference, a4paper]{IEEEtran}

\newif\ifISIT

\ISITfalse 

\usepackage[utf8]{inputenc} 
\usepackage[T1]{fontenc}
\usepackage{url}              
\usepackage{cite}             

\usepackage[cmex10]{amsmath}  
\usepackage{mleftright}       
\mleftright                   

\usepackage{graphicx}         
\usepackage{booktabs}         

\usepackage{algorithm, algorithmic}

\usepackage{color}

\usepackage{amssymb, amsfonts}
\usepackage{bm}
\usepackage{comment}
\usepackage{enumerate}

\newtheorem{lemma}{Lemma}
\newtheorem{theorem}{Theorem}

\newtheorem{definition}{Definition}
\newtheorem{property}{Property}
\newtheorem{remark}{Remark}

\begin{document}

\ifISIT
\title{Computation of Marton's Error Exponent for Discrete Memoryless Sources} 
\else
\title{Computing the optimal error exponential function for fixed-length lossy coding in discrete memoryless sources}
\fi


%
\author{%
   \IEEEauthorblockN{Yutaka Jitsumatsu}
   \IEEEauthorblockA{Tokyo Institute of Technology\\
                     Ookayama 2-12-1, Meguro-ku, Tokyo\\
                     jitsumatsu@ict.e.titech.ac.jp}
                   }

\maketitle

\begin{abstract}
The error exponent of fixed-length lossy source coding was established by Marton. Ahlswede showed that this exponent can be discontinuous at a rate $R$, depending on the probability distribution $P$ of the given information source and the distortion measure $d(x,y)$. The reason for the discontinuity in the error exponent is that there exists $(d,\Delta)$ such that the rate-distortion function $R(\Delta|P)$ is neither concave nor quasi-concave with respect to $P$. Arimoto's algorithm for computing the error exponent in lossy source coding is based on Blahut's parametric representation of the error exponent. However, Blahut's parametric representation is a lower convex envelope of Marton's exponent, and the two do not generally agree. The contribution of this paper is to provide a parametric representation that perfectly matches with the inverse function of Marton's exponent, thus avoiding the problem of the rate-distortion function being non-convex with respect to $P$. The optimal distribution for fixed parameters can be obtained using Arimoto's algorithm. Performing a nonconvex optimization over the parameters successfully yields the inverse function of Marton's exponent.
\end{abstract}

\section{Introduction}
The rate distortion function for 
an independent binary source $P(0) = p$ and $P(1) = 1-p$ with Hamming distortion measure 
is given by~\cite[Chapter 10.3]{CoverTEXT}
\begin{align}
R(\Delta | P )
=
\begin{cases}
h(p) - h(\Delta), & 0\leq \Delta \leq \min \{ p, 1-p \}, \\
0, & \Delta >\min \{ p, 1-p \},
\end{cases}
\end{align}
where $h(p)=-p\log p - (1-p) \log (1-p)$ is a binary entropy function\footnote{In this paper, $\log$ denotes the natural logarithm. }. 
Because $R(\Delta |P)$ of this example is quasi-concave\footnote{
A function $f$ on $\mathcal{P(X)}$ is said to be quasi-convex if for all real $\alpha \in \mathbb{R}$, the set $\{ P \in \mathcal{P(X)}: 
f(x)<\alpha\}$ is convex. A function $f$ is quasi-concave if $-f$ is quasi-convex.
} in $P$, one would expect that it is so in general.
In~\cite{Ahlswede1990}, Ahlswede disproved this conjecture by giving a counterexample that for a fixed $\Delta$, $R(\Delta |P)$ has a local maximum that is different from the global maximum. He showed, as a consequence of this fact, that Marton's optimal error exponent~\cite{Marton1974} can be discontinuous at some rate $R$ for a fixed $\Delta$ and $P$.


For a given information source, the rate distortion function is usually not explicitly expressed, and is defined as the solution to a certain optimization problem. An algorithm for elegantly solving this optimization problem is given by Blahut~\cite{Blahut1972} and, together with Arimoto's algorithm~\cite{Arimoto1972} for computing the channel capacity of a discrete memoryless channel, is called the Arimoto-Blahut algorithm.
Arimoto also gave an algorithm for computing the error exponent for lossy source coding~\cite{Arimoto1976}, but his algorithm is based on Blahut's suboptimal error exponent. 
Marton's exponent is defined as a nonconvex optimization problem, and nonconvex problems often do not have efficient algorithms to solve them.
The computation of Marton's function has been an open problem since Arimoto stated it in~\cite{Arimoto1976}.

The main contribution of this paper is that we establish a parametric expression with two parameters that perfectly matches the inverse function of Marton's error exponent. 
When the parameters are fixed, such an expression involves only convex optimization, which can be computed efficiently by the Arimoto algorithm~\cite{Arimoto1976}. This implies that a non-convex optimization over probability distributions is transformed into a non-convex optimization over two parameters with a convex optimization over probability distributions. 
Using Ahlswede's counterexample, we show that the parametric expression allows to correctly draw the inverse function of Marton's exponent. 

\section{The error exponent for lossy source coding}
We begin with mathematical definitions of the rate distortion function and error exponent of fixed-length lossy source coding. 
Consider a Discrete Memoryless Source (DMS) with 
a source alphabet $\mathcal{X}$ and a reconstruction alphabet $\mathcal{Y}$.
Assume $\mathcal{X}$ and $\mathcal{Y}$ are finite sets.  
The set of probability distributions on $\mathcal{X}$ is denoted by $\mathcal{P(X)}$.
Fix a probability distribution on $\mathcal{X}$, denoted by $P \in \mathcal{P(X)}$. 
Denote a letter-wise distortion measure by $d(x,y) \ge 0$. 
Then, the rate distortion function is given by 
\begin{align}
R(\Delta|P) = 
\min_{
\genfrac{}{}{0pt}{}{
q_{Y|X}\in \mathcal{P(Y|X)}:
}{
\mathrm{E}[d(X,Y) ] \leq \Delta 
}
} I(P, q_{Y|X} ), 
\label{Rate_distortion_function}
\end{align}
where $I(P, q_{Y|X})$ is the mutual information, $\mathcal{P(Y|X)}$ is the set of conditional probability distributions on $\mathcal{Y}$ given $\mathcal{X}$.
Here the expectation of $d(X,Y)$ is taken over the joint probability distributions
$P(x) \cdot q_{Y|X}(y|x)$. We have $R(\Delta |P) = 0$ if $\Delta \geq \Delta_{\max} := 
{\min_y} \sum_{x\in \mathcal{X}} P(x) d(x,y)$.

Marton proved that the following function is the optimal error exponent~\cite{Marton1974}. 
For a fixed $\Delta \in [0,\Delta_{\max}]$, her exponent is defined by 
\begin{align}
E_{\rm M}(R | \Delta, P) =
\min_{
\genfrac{}{}{0pt}{}
  {
    q_X \in \mathcal{P(X)} :
  }{
    R(\Delta | q_X) \geq R 
  }
}
D(q_X || P)
\label{E_M}
\end{align}
for $0\leq R \leq R_{\max}(\Delta) := \max_{ q_X\in \mathcal{P(X)}} R(\Delta | q_X)$,
where $D(\cdot||\cdot)$ denotes the relative entropy. 
From its definition, it is clear that $E_{\rm M}(R | \Delta, P)$ satisfies the following properties.
\begin{property}
\begin{enumerate}[a)]
\item $E_{\rm M}(R | \Delta, P) = 0 $ if $R \leq R(\Delta | P) $.
\item For fixed $\Delta\geq 0$ and $P\in \mathcal{P(X)}$, $E_{\rm M}(R | \Delta, P) $ is a monotone non-decreasing function of $R\in [0, R_{\max} (\Delta) ]$.
\end{enumerate}
\end{property}

Arimoto's computation algorithm for error exponent~\cite{Arimoto1976} is based on 
the parametric expression of Blahut's exponent~\cite{Blahut1974}, defined by 
\begin{align}
&E_{\rm B}(R | \Delta, P)\notag \\
&=
\sup_{\rho\geq 0} 
\bigg\{ \rho R - 
\sup_{\nu\geq 0} 
\Big[
\max_{p_Y} E_{0,\rm s}^{(\rho,\nu)} ( p_Y | P ) 
-
\rho \nu \Delta 
\Big]
\bigg\} \label{def:E_B}
\end{align}
for $0\leq \Delta \leq \Delta_{\rm max}$ and $0\leq R\leq R_{\max}(\Delta)$,
where
\begin{align}
& E_{0,\rm s}^{(\rho,\nu)} ( p_Y | P ) \notag \\
& =
- \log \sum_{x \in \mathcal{P(X)} }
P(x) 
\bigg\{
\sum_{y\in \mathcal{Y}} p_Y(y) {\rm e}^{-\nu d(x,y) }
\bigg\}^{-\rho}.
\label{E_0s}
\end{align}

From Eq.~(\ref{def:E_B}), we can easily see that  
$\rho R -
\sup_{\nu\geq 0} 
\Big[
\max_{p_Y} E_{0,\rm s}^{(\rho,\nu)} ( p_Y | P )
-\rho \nu \Delta 
\Big]$ is the supporting line to the curve $E_{\rm B}(R|\Delta,P)$ with slope $\rho$
and thus $E_{\rm B}(R|\Delta,P)$ is a convex function of $R$. 

\begin{remark} \label{remark3}
In the expression of Blahut's exponent (\ref{def:E_B}), 
$ \max_{p_Y} E_{0,\rm s}^{(\rho,\nu)} ( p_Y | P ) $ 
is not necessarily concave in $\nu$. 
Hence, the computation of Blahut's exponent requires nonlinear optimization over $\nu$. 
\ifISIT
See~\cite{YutakaISIT2023arXiv} for a graph of an example of $\max_{p_Y} E_{0,\rm s}^{(\rho,\nu)}(p_Y|P) - \rho \nu \Delta$ with two local maxima. 
\else
See Appendix~\ref{appendix_graph} for a graph of an example of $\max_{p_Y} E_{0,\rm s}^{(\rho,\nu)}(p_Y|P) - \rho \nu \Delta$ with two local maxima. 
\fi 
\end{remark}

The relation between $E_{\rm M}(R|\Delta,P)$ and
$E_{\rm B}(R|\Delta,P)$ is stated as follows:
\begin{lemma}\label{lemma.1}
For any $P\in \mathcal{P(X)}$, distortion measure $d(x,y)$, $R\geq 0$, and $\Delta\geq0$,
$E_{\rm B}(R|\Delta,P)$ 
is a lower convex envelope of $E_{\rm M}(R|\Delta,P)$.
\end{lemma}

The proof of Lemma~\ref{lemma.1} can be found in~\cite{ArikanMerhav1998} in the context of guessing exponent.
\ifISIT
See also \cite{YutakaISIT2023arXiv}.
\else
To make this paper self-contained, we give the proof in Appendix\ref{appendix_proof}. 
\fi 

To the best of the author's knowledge, any computation method for Marton's error exponent has not been established.
The reason why it is difficult to derive an algorithm for computing Marton's exponent is that 
$R(\Delta|P)$ is not necessarily concave with respect to (w.r.t.) $P$.

Marton's exponent (\ref{E_M}) is rephrased in a standard form of the optimization problem as
\begin{align}
\text{minimize  \,}\quad & D(q_X||P) \label{objective}\\ 
\text{subject to }\quad  & R(\Delta | q_X )\geq R,\\
& q_X(x)\geq 0, \label{condition1}\\
& \sum_{x\in \mathcal{X}} q_X(x) = 1. \label{condition2}
\end{align}
The correct approach to the optimization problem is to find a solution that satisfies the 
Karush–Kuhn–Tucker (KKT) condition and consider the Lagrangian function. 
To do this, we need to evaluate the derivative of $R(\Delta|q_X)$ w.r.t. $q_X$.
Because $R(\Delta|q_X)$ is defined by a constrained optimization problem (\ref{Rate_distortion_function}), another Lagrangian is introduced. 
The author 
was unable to derive a parametric formula that is in exact agreement with Marton's formula. 
We will take a different approach to compute Marton's exponent in Section III.

\section{Main Result}
For a given distortion measure $d(x,y)$, the feasible region $\{ q_X\in \mathcal{P(X)}\colon R(\Delta|q_X)\geq R\}$ in (\ref{E_M}) is not necessarily convex. 
In this case, the computation of Marton's exponent is not easy except for some special cases.
The main contribution of this paper is the establishment of the computation method for Marton's exponent.
Its derivation consists of four steps.

\subsubsection{Inverse function}
The first step is not to find Marton's exponent directly, but first to find its inverse function.
We define the following function. 
\begin{definition}
For $E\geq 0$ and $\Delta\geq 0$, we define 
\begin{align}
R_{\rm M}(E | \Delta, P )
=
\max_{
\genfrac{}{}{0pt}{}{
q_X \in \mathcal{P(X)}:
}{
D(q_X||P) \leq E
}
} R(\Delta | q_X).  \label{R_M}
\end{align}
\end{definition}

The idea of analyzing the inverse function of the error exponent was first introduced by Haroutunian et al.~\cite{Haroutunian1984,Harutyunyan2004}. They defined the rate-reliability-distortion function as the minimum rate at which the messages of a source can be encoded and then reconstructed by the decoder with an exponentially decreasing probability of error, and proved that the optimal rate-reliability-distortion function is given by (\ref{R_M}).

It is clear from the definition that this function satisfies the following basic properties
\begin{property}
\begin{enumerate}[a)]
\item $R_{\rm M}(E | \Delta , P )$ is a monotone non-decreasing function of $E$ for fixed $\Delta\geq 0$
and $P$.
\item $R_{\rm M}(0 | \Delta , P ) = R(\Delta|P)$ holds. 
\item $R_{\rm M}(E | \Delta , P ) = R_{\max}(\Delta)$ for $E \geq  D(q_X^*||P)$, where
$q_X^* = \arg \max_{q_X} R(\Delta|q_X)$. 
\end{enumerate}
\end{property}

\subsubsection{A parametric expression for the rate distortion function}
The function $R_{\rm M}(E|\Delta, P)$ is much easier to analyze than (\ref{E_M}) because 
the feasible region for the maximization in (\ref{R_M}) is convex.
In (\ref{R_M}), however, the objective function is the rate distortion function, which is not necessarily convex.
To circumvent this issue, we use the following parametric expression of $R(\Delta|q_X)$. 
This is the second step. 
\begin{lemma}
\label{lemma4}
We have
\begin{align}
& R(\Delta | q_X) =
\sup_{\nu \geq 0}
\Big[ -\nu \Delta \notag\\
&\hspace{8mm}  
+ \min_{ p_Y \in \mathcal{P(Y)} }
-\sum_x q_X(x) \log \sum_y p_Y(y) \mathrm{e}^{-\nu d(x,y)}
\Big]. \label{parametric_Rate_distortion}
\end{align}
\end{lemma}

One can refer~\cite[Corollary 8.5]{Csiszar-KornerBook} for the proof. 
\ifISIT
\else
However, to make this paper self-contained we give the proof in Appendix\ref{appendix_proof}. 
\fi

We should mention that the expression (\ref{parametric_Rate_distortion}) is related to an important notion of $d$-tilted information density~\cite{Kostina2012}, although this relation is not used in this paper. 
Denote the $\nu$ and $p_Y$ that attains (\ref{parametric_Rate_distortion}) by $\nu^*$ and $p_Y^*$.
Then, 
\begin{align}
\jmath_X(x,d) := -\log \sum_y p_Y^*(y) {\rm e}^{-\nu^* (d(x,y)-\Delta)}
\end{align}
is called $d$-tilted information and we observe that 
$R(\Delta|q_X) = {\rm E}_{q_X} [ \jmath_X(X,d) ]$ holds.

\subsubsection{Minimax theorem}
We substitute (\ref{parametric_Rate_distortion}) into (\ref{R_M}). 
Then, except for the maximization over $\nu\ge 0$, we have to evaluate the following saddle point w.r.t. two probability distributions: 
    \begin{align}
    \max_{
\genfrac{}{}{0pt}{}{
q_X \in \mathcal{P(X)}:
}{
D(q_X||P) \leq E
}
} 
\min_{ p_Y \in \mathcal{P(Y)} }
-\sum_x q_X(x) \log \sum_y p_Y(y) \mathrm{e}^{-\nu d(x,y)}    
    \label{saddlepoint}
    \end{align}

The third step is the exchange of the order of max and min
in (\ref{saddlepoint}). 
For deriving an algorithm for computing $R_{\rm M}(E|\Delta,P)$, 
the saddle point (\ref{saddlepoint}) should be transformed into minimization or maximization problems. 
In order to derive such an expression, we exchange of the order of maximization w.r.t.~$q_X$ and minimization w.r.t.~$p_Y$.
The following lemma is essential for deriving the exact parametric expression for the inverse function of the error exponent.
\begin{lemma} \label{application_of_minimax_theorem}
    For any $E\ge 0$ and $\nu\ge 0$, we have
    \begin{align}
    & 
    \max_{
    \genfrac{}{}{0pt}{}{ q_X \in \mathcal{P(X)}: } 
    { D(q_X||P) \leq E }
    } 
\min_{ p_Y \in \mathcal{P(Y)} }
-\sum_x q_X(x) \log \sum_y p_Y(y) \mathrm{e}^{-\nu d(x,y)}
 \notag\\
    &=
    \min_{p_Y \in \mathcal{P(Y)}} 
    \max_{
    \genfrac{}{}{0pt}{}{q_X \in \mathcal{P(X)}: }{ D(q_X||P) \leq E } 
    }
    -\sum_x q_X(x) \log \sum_y p_Y(y) {\rm e}^{-\nu d(x,y) } 
    \end{align}
\end{lemma}

The validity of this exchange relies on Sion's minimax theorem~\cite{Sion1958}.

\begin{theorem}[Sion~\cite{Sion1958}] \label{MinimaxTheorem}
Let $\mathcal{P}$ and $\mathcal{Q}$ be convex, compact spaces, and
$f(p, q)$ a function on $\mathcal{P\times Q}$. 
If $f(p, q)$ is lower semicontinuous and quasi-convex on $p \in \mathcal{P}$
for any fixed $q\in\mathcal{Q}$
and $f(p, q)$ is upper semicontinuous and quasi-concave in $q \in \mathcal{Q}$
for any fixed $p\in\mathcal{P}$, then
\begin{align}
    \inf_{ p \in \mathcal{P} } \sup_{ q \in \mathcal{Q} } f(p,q) = 
    \sup_{ q \in \mathcal{Q} } \inf_{ p \in \mathcal{P} } f(p,q) .
\end{align}
\end{theorem}

\textit{Proof of Lemma~\ref{application_of_minimax_theorem}}: 
As stated above, the objective function of (\ref{saddlepoint}) is linear in $q_X$ and convex in $p_Y$. 
Hence, we can apply Theorem~\ref{MinimaxTheorem} to (\ref{saddlepoint}). 
A direct application of Theorem~\ref{MinimaxTheorem}
proves Lemma~\ref{application_of_minimax_theorem}.
\hfill\IEEEQED

\subsubsection{The second Lagrange multiplier}
Next, we define the following functions:
\begin{definition}
\label{def:Gmunu}
For $\mu\geq0, \nu\geq0$, $p_Y\in \mathcal{P(Y)}$,
and $P\in \mathcal{P(X)}$, we define 
\begin{align}
&G^{(\nu)}(E, p_Y |P)\notag\\
&= 
    \max_{
    \genfrac{}{}{0pt}{}{q_X \in \mathcal{P(X)}: }{ D(q_X||P) \leq E } 
    }
    -\sum_x q_X(x) \log \sum_y p_Y(y) {\rm e}^{-\nu d(x,y) } 
    \label{GnuEp_YP}
    \\
&G^{(\mu,\nu)}(p_Y|P) 
= \max_{q_X\in \mathcal{P(Y)}} \bigg[
- \mu D( q_X || P )
\notag\\
&\hspace{8mm}
-\sum_{x} q_X(x) \log \sum_y p_Y(y) \mathrm{e}^{-\nu d(x,y)} \bigg], 
\label{Gmunup_YP}
\\
& G^{(\mu, \nu)}(P) = \min_{p_Y \in \mathcal{P(Y)}} G^{(\mu, \nu)}(p_Y|P). 
\label{GmunuP}
\end{align}
\end{definition}

The last step is to transform (\ref{GnuEp_YP}), which is a constrained maximization, into an unconstrained maximization by  introducing a Lagrange multiplier. For this purpose, we have defined (\ref{Gmunup_YP}). Then,  (\ref{Gmunup_YP}) is explicitly obtained as follows:
\begin{lemma}
    \label{Lemma_G_nu}
For $\mu,\nu\geq 0$, $p_Y\in \mathcal{P(Y)}$,
and $P\in \mathcal{P(X)}$, we have
\begin{align}
&
G^{(\mu,\nu)} ( p_Y | P) \notag\\
&=
\begin{cases}
\mu \log \sum_x P(x) \left\{ \sum_{y} p_Y (y) \mathrm{e}^{-\nu d(x,y)} \right\}^{-1/\mu} & \text{ if } \mu > 0,\\
- \log \min_x \sum_y p_Y(y) \mathrm{e}^{-\nu d(x,y)} &\text{ if } \mu = 0.
\end{cases}  
\label{Gmunup_YP2}
\end{align}
\end{lemma}

We have the following lemma.
\begin{lemma}
    \label{lemma3}
    For $\nu\geq 0,E\geq 0, p_Y\in \mathcal{P(Y)}$, and $P\in\mathcal{P(X)}$, we have
\begin{align}
G^{(\nu)}(E, p_Y|P) = \inf_{\mu\ge 0} \{ \mu E + G^{(\mu,\nu)} (p_Y | P) \}. \label{G_nu2}
\end{align}
\end{lemma}
\ifISIT
The proofs of Lemmas~\ref{Lemma_G_nu} and~\ref{lemma3} appear in~\cite{YutakaISIT2023arXiv}.
\else
The proofs of Lemmas~\ref{Lemma_G_nu} and~\ref{lemma3} appear in Section~\ref{Section:Proof}.
\fi

Eq.(\ref{G_nu2}) is a parametric expression for (\ref{GnuEp_YP}).
Finally, we obtain the following theorem.
\begin{theorem}
\label{theorem2}
For any $P\in \mathcal{P(X)}$, $0\leq E \leq E_{\max}$, and $0\leq \Delta \leq \Delta_{\max}$, we have
\begin{align}
R_{\rm M}(E| \Delta, P ) =
\sup_{\nu\geq 0} 
\inf_{\mu \geq 0}
\left[ 
- \nu \Delta + \mu E 
+G^{(\mu, \nu)} ( P )   
\right]. \label{eq.theorem2}
\end{align}
\end{theorem}

\ifISIT 
See~\cite{YutakaISIT2023arXiv} for the proof.
\else 
\textit{Proof:}
We have the following chain of equations.
\begin{align}
    & R_{\rm M}(E|\Delta, P) \notag\\
    & 
    \stackrel{\rm (a)}
    = 
    \max_{
    \genfrac{}{}{0pt}{}{ q_X \in \mathcal{P(X)}: } 
    { D(q_X||P) \leq E }
    }
    \sup_{\nu \geq 0}
\Big[ -\nu \Delta \notag\\
&\hspace{8mm}  
+ 
\min_{ p_Y \in \mathcal{P(Y)} }
-\sum_x q_X(x) \log \sum_y p_Y(y) \mathrm{e}^{-\nu d(x,y)}
\Big] \notag\\
    & 
    \stackrel{\rm (b)}
    = 
    \sup_{\nu \geq 0}
    \min_{ p_Y \in \mathcal{P(Y)} }
    \max_{
    \genfrac{}{}{0pt}{}{ q_X \in \mathcal{P(X)}: } 
    { D(q_X||P) \leq E }
    }
    \Big[ -\nu \Delta \notag\\
&\hspace{8mm} -\sum_x q_X(x) \log \sum_y p_Y(y) \mathrm{e}^{-\nu d(x,y)}
\Big] \notag\\
    & 
    \stackrel{\rm (c)}
    = 
    \sup_{\nu \geq 0}
    \min_{ p_Y \in \mathcal{P(Y)} }
    \Big[ -\nu \Delta + G^{(\nu)}(E, p_Y|P) 
\Big] \notag\\
    &
    \stackrel{\rm (d)}
    = 
    \sup_{ \nu \ge 0 }
    \inf_{ \mu \ge 0 }
    \min_{ p_Y \in \mathcal{P(Y)} }
    \Big[ -\nu \Delta + \mu E + G^{(\mu, \nu)} (q_Y|P)
\Big] \notag\\ 
    &
    \stackrel{\rm (e)}
    = 
    \sup_{ \nu \ge 0 }
    \inf_{ \mu \ge 0 }
    \Big[ -\nu \Delta + \mu E + G^{(\mu, \nu)} (P)
\Big] 
\end{align}
Step (a) follows from Lemma~\ref{lemma4},
Step (b) follows from Lemma~\ref{application_of_minimax_theorem},
Step (c) follows from Eq.(\ref{GnuEp_YP}),  
Step (d) follows from Lemma~\ref{lemma3}, 
and Step (e) follows from Eq.(\ref{GmunuP}). \hfill\IEEEQED
\fi

Eq.~(\ref{eq.theorem2}) is valuable because it is an equation that is in perfect agreement with the inverse function of Marton's optimal error exponent. Such an exact parametric expression has not been known before.

Note that $ G^{(\mu, \nu)} (p_Y|P)$ for $\mu>0$ in (\ref{Gmunup_YP2}) is equal to (\ref{E_0s}) with $\rho = 1/\mu$ multiplied by $-\mu$.
Therefore, $\min_{p_Y} G^{(\mu, \nu)} (p_Y|P)$  is computed by Arimoto's algorithm~\cite{Arimoto1976} with $\rho=1/\mu$ if $\mu>0$. If $\mu=0$, minimization of $G^{(\mu,\nu)}(p_Y|P)$ reduces to 
a linear programming problem. Our proposed method is stated as follows:

{\bf [Proposed Method for computing $ R_{\rm M}(E,\Delta | P )$]} 
\begin{enumerate}
\item 
Set $\mu_i = i  (\Delta\mu)$, $\nu_j = j (\Delta \nu)$, and $E_k = k (\Delta E)$ for $i=0,1,\ldots, N-1$,
$j=0,1,\ldots, M-1$, and $k=0,1,\ldots, K-1$, where $N$, $M$, $K$, $(\Delta\mu)$, $(\Delta \nu)$, and $(\Delta E)$ 
are determined beforehand according to the precision.
\item 
For each $i$ and $j$, compute $G^{(\mu_i, \nu_j)} ( P) $.
For $i>0$, this is computed by Arimoto algorithm with $\rho = 1/\mu_i$.
Arimoto algorithm is shown in Algorithm~\ref{algorithm1}.
If $i=0$, solve the linear programming problem:
\begin{align}
\text{maximize} \quad  & c \\
\text{subject to} \quad &\sum_y p_Y(y) \mathrm{e}^{-\nu_j d(x,y)} \ge c, \quad ^\forall x\in\mathcal{X}
\end{align}
with (\ref{condition1}) and (\ref{condition2}), where variables are $q_Y(y)$ and $c$.
Obtain the maximum value $c=c^*$ and we have $G^{(0, \nu_j)}(P) = -\log c^*$.
\item Let $a_{j,k}  = \min_{i} \{ G^{(\mu_i, \nu_j)}(P) + \mu_i E_k \} $.
\item Finally, $R_{\rm M}(E_k|\Delta, P) = \max_j \{ a_{j,k} - \nu_j \Delta\}$ is obtained.
\end{enumerate}

\begin{remark}
Since $G^{(\mu,\nu)}(P)$ lacks the convex property, the grid-based brute-force optimization is a reasonable choice. 
We must emphasize the fact that before this paper, we had no efficient way to compute Marton's exponent. 
The brute-force computational cost for the optimization problem of (\ref{objective})-(\ref{condition2}) is exponential in $|\mathcal{X}|$.
Compared to this, the computational cost for the two-dimensional search is not significant. 
\end{remark}

\ifISIT
\else 
\begin{remark}
Minimizing over $\mu$ and maximizing over $\nu$ must be done in this order, as defined in (\ref{eq.theorem2}).
Interchanging the order of the operations yields 
\begin{align}
&R_{\rm M}(E| \Delta, P ) \notag \\
&\leq 
\inf_{\mu \geq 0}
\left\{ \sup_{\nu\geq 0} 
\left[
\min_{p_Y} G^{(\mu, \nu)} ( p_Y | P)
- \nu \Delta 
\right ]
+ \mu E 
\right\}\notag\\
& =:
\tilde R_{\rm M}(E| \Delta, P ),    
\end{align}
which is concave in $E$ and does not match with (\ref{eq.theorem2}) in general. 
\end{remark}
\fi

\begin{algorithm}[t]
  \begin{algorithmic}[0]
  \caption{Arimoto algorithm for computing the error exponent of lossy source coding~\cite{Arimoto1976}}
  \label{algorithm1}
  \STATE{
  This algorithm includes AB algorithm for the rate distortion function~\cite{Blahut1972} as a special case of $\rho = 0$.}
  \REQUIRE{$\nu, \rho\geq 0$, $d$, and $P$ are given.}
  \STATE{Choose initial output distribution $p_Y^{[0]}$ arbitrarily so that all elements are nonzero.}
  \FOR{$t=0,1,2,\ldots$}
  \STATE \begin{align}
  &\hspace{-6mm} q_{Y|X}^{[t]}(y|x) = \frac{ p_Y(y) \mathrm{e}^{-\nu d(x,y)} }{ \sum_{y\in \mathcal{Y} } p_Y^{[t]}(y) \mathrm{e}^{-\nu d(x,y)}} \\
  &\hspace{-6mm}p_Y^{[t+1]}(y) = 
  \frac{ \Big[ {\displaystyle \sum_x }P(x) {\rm e}^{\rho\nu d(x,y) } q_{Y|X}^{[i]}(y|x)^{1+\rho} 
  \Big]^{\frac{1}{1+\rho} } }
  { { \displaystyle \sum_{y'}}
  \Big[ {\displaystyle \sum_x }P(x) {\rm e}^{\rho\nu d(x,y') } q_{Y|X}^{[i]}(y'|x)^{1+\rho} 
  \Big]^{\frac{1}{1+\rho} } 
  }
  \end{align}
  \ENDFOR
  \end{algorithmic}
\end{algorithm}

\section{Ahlswede's Counterexample}
\label{sec:counterexample}
%
The discussion about the continuity of Marton's function was settled by Ahlswede~\cite{Ahlswede1990}.  
In this section, using his counterexample, we show the case where $E_{\rm M}(R, \Delta|P)$ is discontinuous at an $R$.

Ahlswede's counterexample is defined as follows:
Let $\mathcal{Y}=\mathcal{X}$ and $\mathcal{X}$ is partitioned into $\mathcal{X}_A$ and $ \mathcal{X}_B$. 
Define the distortion measure as
\begin{align}
d(x,y) 
= 
\begin{cases}
  0, \text{ if } x=y \in \mathcal{X}, \\
  1, \text{ if } x\neq y \text{ and } x, y \in \mathcal{X}_A,\\
  a, \text{ if } x\neq y \text{ and } x, y \in \mathcal{X}_B,\\
  b, \text{ otherwise. } 
\end{cases}
\label{distortion}
\end{align}
The constant $b$ is sufficiently large value so that encoding a source output $x\in \mathcal{X}_A$ into $y \in \mathcal{X}_B$ or vise versa has a large penalty. 
The constant $a$ is determined later.
We see that distortion measure (\ref{distortion}) is not a strange situation but
can match a situation that we must distinguish whether $x$ is in $\mathcal{X}_A$ or $\mathcal{X}_B$ nearly perfectly.

Assume $|\mathcal{X}_B| = | \mathcal{X}_A|^3$, 
where $|\cdot |$ denotes the cardinality of a set.
Let $Q_A$ and $Q_B$ be uniform distributions on $\mathcal{X}_A$ and $\mathcal{X}_B$, that is, 
\begin{align}
Q_A(x) &= \begin{cases}
1/|\mathcal{X}_A|, &\text{ if } x \in \mathcal{X}_A, \\
0, & \text{ if } x \in \mathcal{X}_B,
\end{cases} \\
Q_B(x) &= \begin{cases}
0, &\text{ if } x \in \mathcal{X}_A, \\
1/|\mathcal{X}_B|, & \text{ if } x \in \mathcal{X}_B. 
\end{cases}
\end{align}
For $\lambda\in [0,1]$, we denote $Q_\lambda = \lambda Q_A+(1-\lambda)Q_B$.
The rate distortion function of $Q_A$ and $Q_B$ are 
\begin{align}
R(\Delta | Q_A) &= \log | \mathcal{X}_A | - h(\Delta) - \Delta \log ( | \mathcal{X}_A | - 1 ),\\ 
R(\Delta | Q_B) &= \log | \mathcal{X}_B | - h( \textstyle \frac{\Delta}{a}  ) - \frac{\Delta}{a}  \log ( | \mathcal{X}_B | - 1 ). 
\end{align}
To simplify the calculation, Ahlswede chose 
the parameters $a$ and $\Delta$ so that 
\begin{align}
\textstyle \frac{\Delta}{a} & =  1 - \Delta, \label{def_a} 
\end{align}
\begin{align}
& \log | \mathcal{X}_A | - \Delta \log ( | \mathcal{X}_A | - 1 )\notag\\
& =
\log | \mathcal{X}_B | - (1-\Delta)  \log ( | \mathcal{X}_B | - 1 )
\label{def_Delta}
\end{align}
hold.

The conjecture that $R(\Delta|P)$ is quasi-convex in $P$ for any given $d(x,y)$ 
and $\Delta$ is disproved 
if $R(\Delta|P)$ is not quasi-convex on any subset of $\mathcal{P(X)}$ for some $d(x,y)$ and some $\Delta$.  
Using the distortion function (\ref{distortion}) and the parameters $a,\Delta$
determined by (\ref{def_a}), (\ref{def_Delta}), 
Ahlswede analyzed the rate distortion function $R(\Delta|P)$ for 
$P \in \{ Q_\lambda= \lambda Q_A + (1-\lambda) Q_B: 0\leq \lambda\leq 1\} \subset \mathcal{P(X)}$
and showed that 
if $|\mathcal{X}_A|$ is sufficiently large, 
$R(\Delta | Q_{\lambda} ) $ has local maximum different from the global maximum. 
This suggests that $R(\Delta | P ) $ of this case is not quasi-concave in $P$.

\begin{figure}
\centering
\includegraphics[width = 0.75\columnwidth]{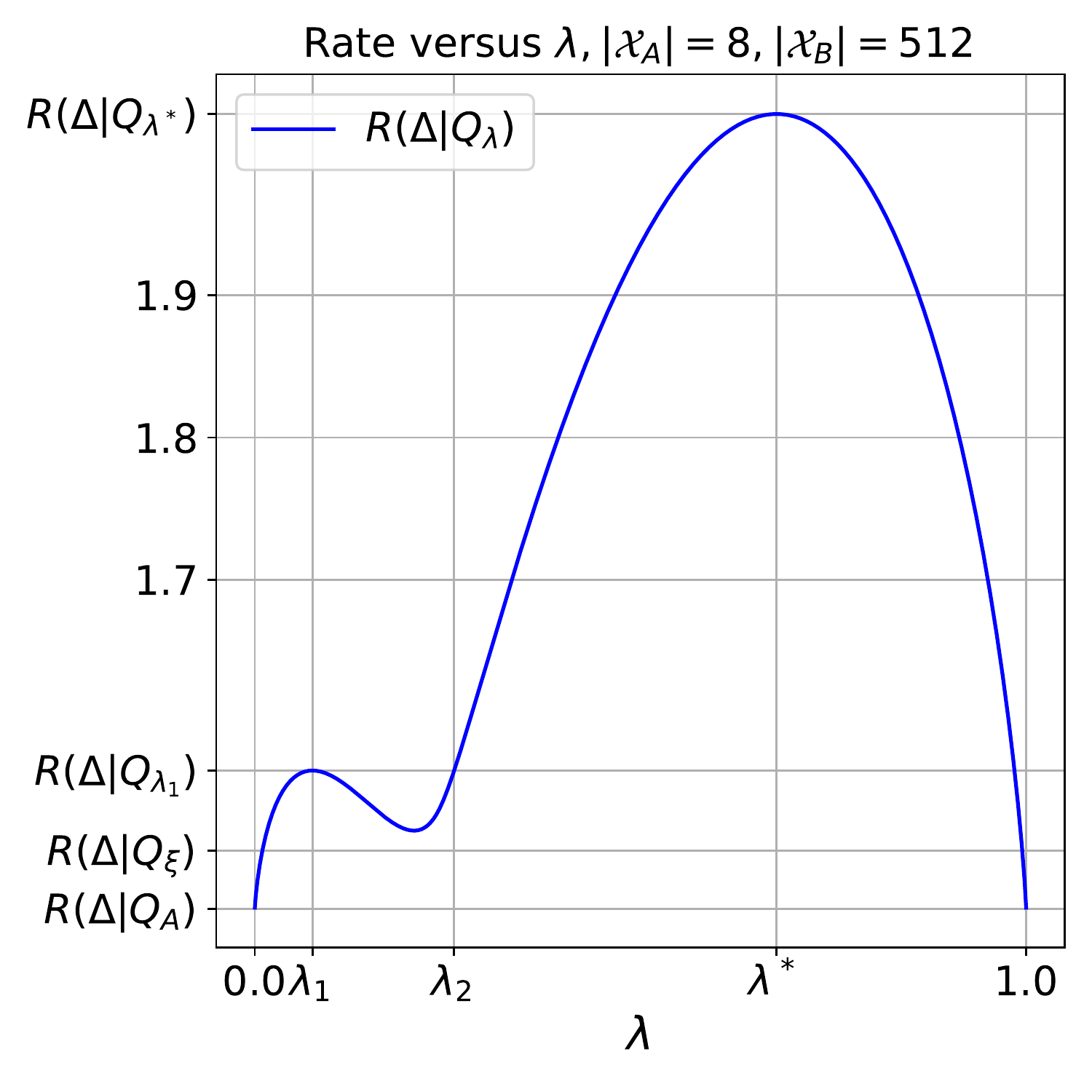}
\caption{Rate distortion function $R(\Delta|Q_{\lambda})$ of Ahlswede's counterexample with $|\mathcal{X}_A|=8$
as a function of $\lambda$. The rate is in unit bit}
\label{fig.1}
\end{figure}

In~\cite{Ahlswede1990}, no graph for $R(\Delta |P)$ was provided. 
We compute the rate distortion function by Arimoto-Blahut algorithm~\cite{Blahut1972, Arimoto1972}
In Fig.~\ref{fig.1}, $R(\Delta |Q_\lambda)$ as a function of $\lambda\in [0,1]$
is illustrated, where $ | \mathcal{X}_A | = 8 $, $\Delta=0.254$, and $a=0.340$. 
If $|\mathcal{X}|$ is smaller than $8$, the graph of $R(\Delta|Q_{\lambda})$ does not have local maximum that is different from the global maximum. 
We observe $R(\Delta |Q_\lambda)$ is bimodal with global maximum at $\lambda = \lambda^* = 0.676$
and local maximum at $\lambda = \lambda_1 = 0.0746$.

Next, let us draw the graph of the error exponent using the rate distortion function in Fig.~\ref{fig.1}.
We give the following theorem to evaluate the error exponent for the Ahlswede's counterexample.
\begin{theorem}
\label{theorem1}
Assume the distortion measure $d(x,y)$ is given by (\ref{distortion}) and
let $P=Q_\xi$ for a fixed $\xi\in [0,1]$.
Then, we have
\begin{align}
E_{\rm M}(R | \Delta, Q_{\xi})
=
\min_{
\genfrac{}{}{0pt}{}{
\lambda \in[0,1]:
}{
R(\Delta |Q_\lambda) \geq R
}
}
D_2(\lambda || \xi )
\end{align}
where $D_2(p||q) = p\log\frac{p}{q} + (1-p) \log \frac{1-p}{1-q}$ is a binary divergence.
\end{theorem}

\ifISIT
See~\cite{YutakaISIT2023arXiv} for the proof.
\else 
Before giving the proof, we state the following lemma due to Ahlswede~\cite{Ahlswede1990}.
\begin{lemma}
\label{lemma1}
For any $P\in \mathcal{P(X)}$ with $\mathcal{X}=\mathcal{X}_A \cup \mathcal{X}_B$ where $\mathcal{X}_A$ and $\mathcal{X_B}$ are disjoint, 
define $\xi=  \sum_{x\in \mathcal{X}_A} P(x) $. We have
\begin{align}
R(\Delta | \xi Q_A + (1-\xi) Q_B ) \geq R(\Delta |P).
\end{align}
\end{lemma}
See~\cite{Ahlswede1990} for the proof.

\textit{Proof of Theorem~\ref{theorem1}:}
Let $q_X^* \in \mathcal{P(X)}$ be an optimal distribution that attains 
$E_{\rm M}(R | \Delta , Q_{\lambda})$. 
Put $\xi = \sum_{x\in \mathcal{X}_A } q_X^*(x)$.
We will show $q_X^*$ is expressed by $\xi Q_A + (1-\xi) Q_B$. 

From Lemma~\ref{lemma1}, we have 
$ R(\Delta | \xi Q_A + (1-\xi) Q_B) \geq R(\Delta | q_X^*) $ ($\geq R$).
Therefore $ \xi Q_A + ( 1-\xi) Q_B$ is feasible. 
Let $q_A^*(x) = q_X^*(x)/\xi$ for $x\in \mathcal{X}_A$ and
$q_B(x)= q_X^*(x)/(1-\xi)$ for $x\in \mathcal{X}_B$.
Then, we have
\begin{align}
& D(q_X^* || Q_{\lambda} ) \notag\\
& =
  \sum_{x\in \mathcal{X} } q_X^{*}(x) \log \frac{q_X^*(x)}{ Q_{\lambda} (x) }  \notag \\
&=
\sum_{x\in \mathcal{X}_A } \xi q_A^{*}(x) \log \frac{ \xi q_A^{*}(x) }{ \frac{ \lambda }{ |\mathcal{X}_A| } }   \notag\\
& \quad + 
\sum_{x\in \mathcal{X}_B } (1-\xi) q_B^{*}(x) \log \frac{ (1-\xi) q_B^*(x)}{ \frac{1-\lambda}{ |\mathcal{X}_B| }  } 
\notag\\
&=
\xi \left \{ \log \frac{ \xi | \mathcal{X}_A | }{\lambda}
+ \sum_{x\in\mathcal{X}_A} q_A^*(x) \log q_A^*(x) \right \} \notag\\
& \quad+
(1-\xi) \left\{  \log \frac{(1-\xi) | \mathcal{X}_B | }{1-\lambda}
+ \sum_{x\in\mathcal{X}_B} q_B^*(x) \log q_B^*(x) \right\} \notag\\
&
\stackrel{(a)}
\geq
\xi \left \{ \log \frac{ \xi | \mathcal{X}_A | }{\lambda}
-\log |\mathcal{X}_A| \right \} \notag\\
&\quad +
(1-\xi) \left\{  \log \frac{(1-\xi) | \mathcal{X}_B | }{1-\lambda}
-\log |\mathcal{X}_B| \right\} \notag\\
&=D_2(\xi || \lambda) = D ( \xi Q_A + ( 1-\xi) Q_B || Q_{\lambda} ),
\label{inequality}
\end{align}
Equality in (a) holds if and only if $q_A^*(x) = 1/|\mathcal{X}_A|$
 and $q_B^*(x) = 1/|\mathcal{X}_B|$. 
Since we assumed $q_X^*$ is optimal, we must have $q_X^* = \xi Q_A + ( 1-\xi) Q_B$. 
This completes the proof.\hfill$\IEEEQED$
\fi

Theorem~\ref{theorem1} ensures that the optimal error exponent can be computed as follows:

{\bf [Computation method of the error exponent for Ahlswede's counterexample]}

Let $N$ be a large positive integer and let $\lambda_i = i/N$ for $i=0,1,\ldots, N$.
Compute $R_i = R(\Delta|Q_{\lambda_i})$ and $D_i = D_2(\lambda_i || \xi )$.
Then, arrange $(R_i, D_i)$ in ascending order of $R_i$.
Put $E_i = \min_{j\geq i} D_j$.
Then, by plotting $(R_i, E_i)$ for $i=0,1,\ldots, N$, we obtain
the graph of $E=E_{\rm M}(R| \Delta, Q_{\xi})$ for $R(\Delta|Q_\xi)\leq R \leq R_{\max}$.
We can add a straight line segment $E=0$ for $0\leq R \leq R(\Delta|Q_\xi)$.

\begin{figure}
\centering
\includegraphics[width=0.76\columnwidth]{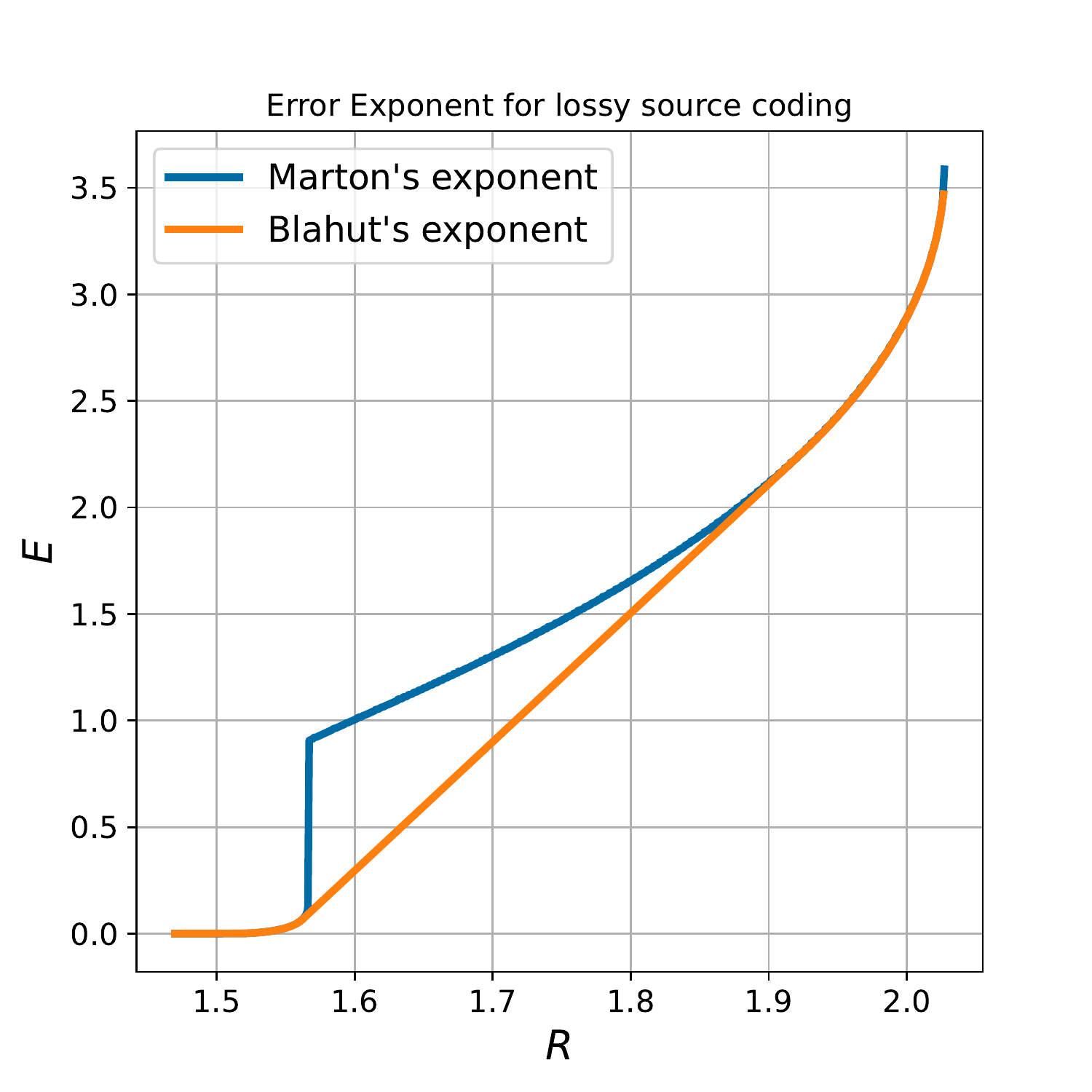}
\caption{Marton's and Blahut's error exponents are illustrated 
as functions of $R$ for Ahlswede's counterexample of Fig.~\ref{fig.1},
where $\Delta=0.254$ and $P=Q_{0.01}$.
}
\label{fig.error_exponent}
\end{figure}
Fig.~\ref{fig.error_exponent} shows the error exponent for Ahlswede's counterexample of Fig.~\ref{fig.1}.
The probability distribution of the source is chosen as 
$P=Q_{\xi}$ with $\xi = 0.01$. 
We observe that $E_{\rm M}(R|\Delta, P) =0$ for $R\leq R(\Delta|Q_{0.01}) = 1.510$
and $E_{\rm M}(R|\Delta, P)$ gradually increases for $1.510\leq R \leq R(\Delta|Q_{ \lambda_1})=1.566$. 
At $R=1.566$, the curve jumps from $E = D(Q_{\lambda_1} || Q_{\xi}) =0.126$
to $E = D( Q_{\lambda_2} || Q_\xi) = 0.904$, where 
$\lambda_2 = 0.258$ satisfies $ R(\Delta|Q_{\lambda_1}) = R(\Delta | Q_{\lambda_2})$. 
For $ R(\Delta|Q_{\lambda_1}) < R \leq R(\Delta |Q_{\lambda^*})$,
the graph is expressed by $(R,E) = ( R(\Delta|Q_\lambda), D(Q_\lambda||Q_\xi) )$ with $ \lambda \in (\lambda_2, \lambda^*)$. 

In Fig.~\ref{fig.error_exponent}, Blahut's parametric expression (\ref{def:E_B}) 
of error exponent is also plotted, where optimal distribution $p_Y^*$
for (\ref{def:E_B}) is computed by Algorithm~\ref{algorithm1}.
This figure clearly shows that there is a gap between these two exponents.

\begin{figure}
    \centering
    \includegraphics[width=0.76\columnwidth]{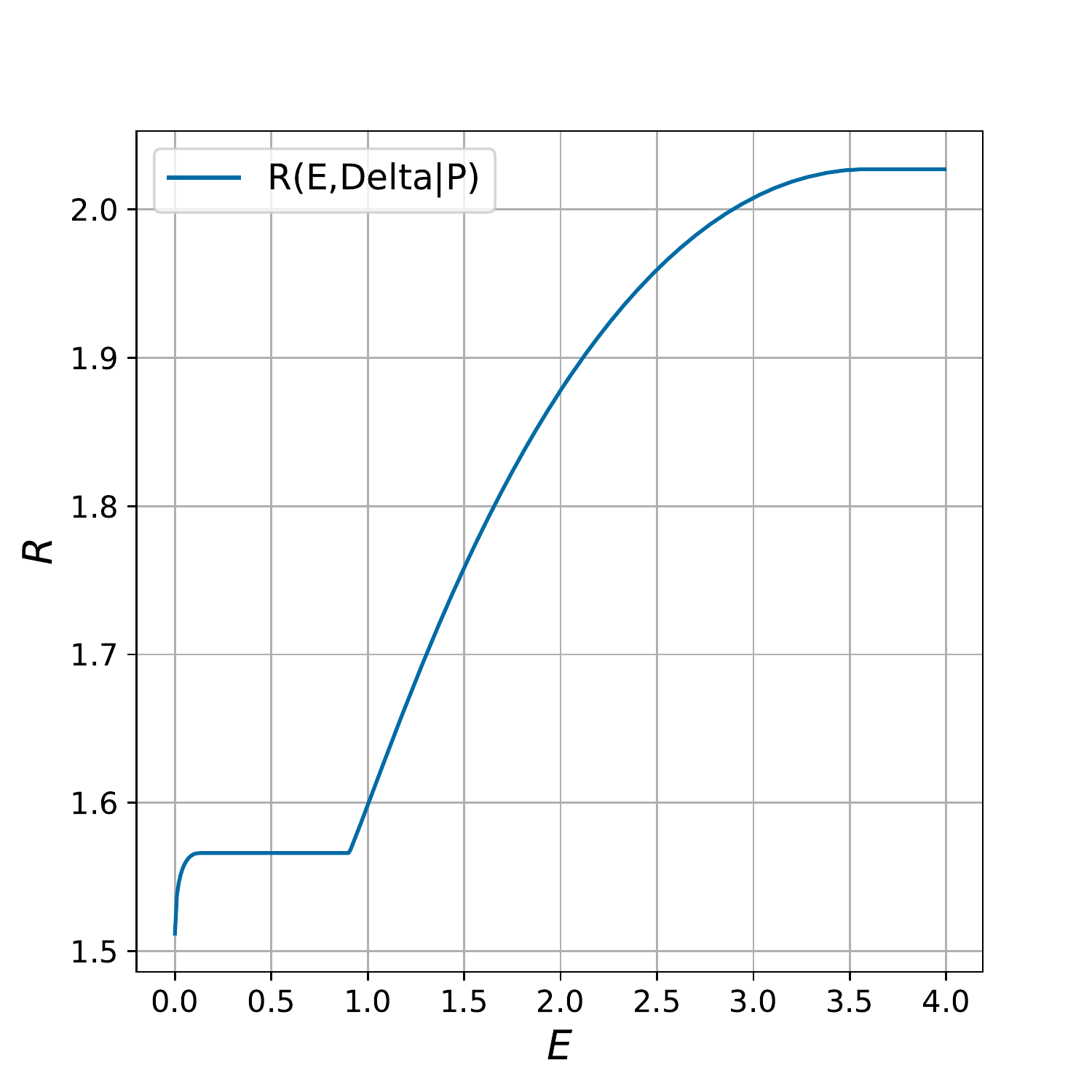}
    \caption{$R_{\rm M}(E | \Delta , P)$ for Ahlswede's counterexample of Fig.~\ref{fig.1}.
    }
    \label{fig.inverse}
\end{figure}

Using the proposed method, we compute $R_{\rm M}(E | \Delta, P)$ for the same parameters for Fig.\ref{fig.error_exponent} by the proposed method. The graph is shown in Fig.~\ref{fig.inverse}. 
It is confirmed that $R_{\rm M}(E|\Delta, P)$ is correctly computed. 
The inverse function is continuous in $E$ and if the inverse function takes a constant value $R_0$ for some finite interval $[E_1, E_2]$, it means the error exponent jumps from $E_1$ to $E_2$ at $R=R_0$.  
Note that while Marton's exponent in Fig~\ref{fig.error_exponent} was computed based on Theorem~\ref{theorem1}, which holds only for Ahlswede's counterexamples, the proposed method is applicable to any $P$, $d$, and $\Delta$.

\ifISIT
\else 
Here is another example to show the discontinuity of the optimal error exponent more clearly.
Let $|\mathcal{X}_A|=50$ and $|\mathcal{X}_B|=|\mathcal{X}_A|^2$ and use the distortion measure  (\ref{distortion}) and determine the parameters $a=0.501$ and $\Delta=0.333$ to satisfy (\ref{def_a}) and (\ref{def_Delta}).
The second example of Marton's error exponent is shown in Fig.~\ref{fig:rate_distortion2}. 
The global maximum is found at $\lambda^* = 0.762$ and a local maximum at $\lambda=\lambda_1=0.065$.
Then, the rate distortion function of this case was computed by Arimoto-Blahut algorithm. 
Marton's exponent and Blahut's error exponents are shown in Fig.~\ref{fig:error_exponent2}, where $P=Q_\xi$ with $\xi = 0.2$. 
We observe that Marton's exponent jumps from $D(Q_{\lambda_1}||Q_{0.2})=0.103$ to $D(Q_{\lambda_2}||Q_{0.2})=0.220$ at $R=R(\Delta|Q_{\lambda_1}) = R(\Delta|Q_{\lambda_2}) = 2.940$. 
In Fig.~\ref{fig:inverse2}, $R_{\rm M}(E|\Delta,P)$ computed by the proposed method is drawn. 
We confirm that the graph is correctly computed.

\begin{figure}
    \centering
    \includegraphics[width=0.8\columnwidth]{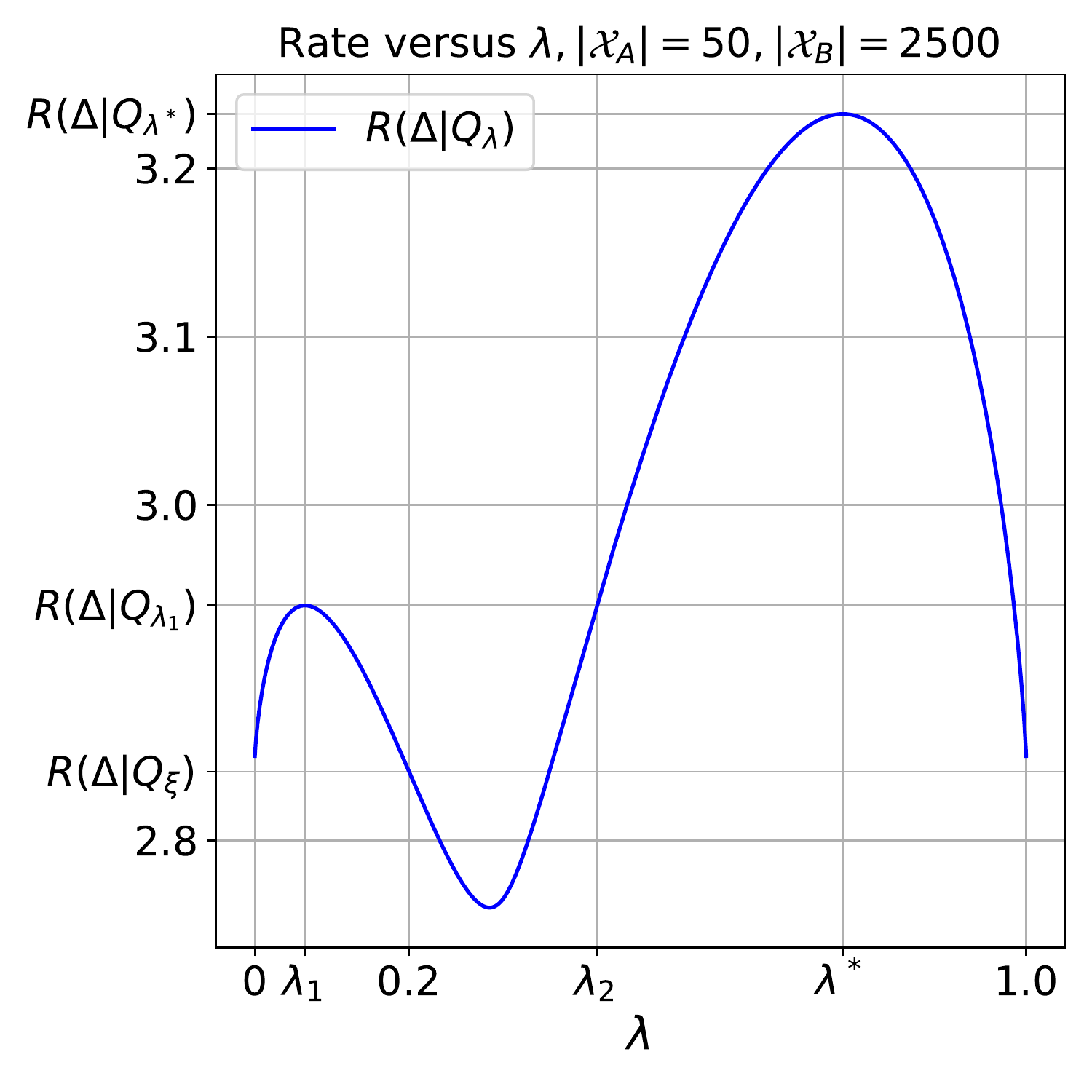}
    \caption{The rate distortion function for the second example}
    \label{fig:rate_distortion2}
\end{figure}

\begin{figure}
    \centering
    \includegraphics[width=0.8\columnwidth]{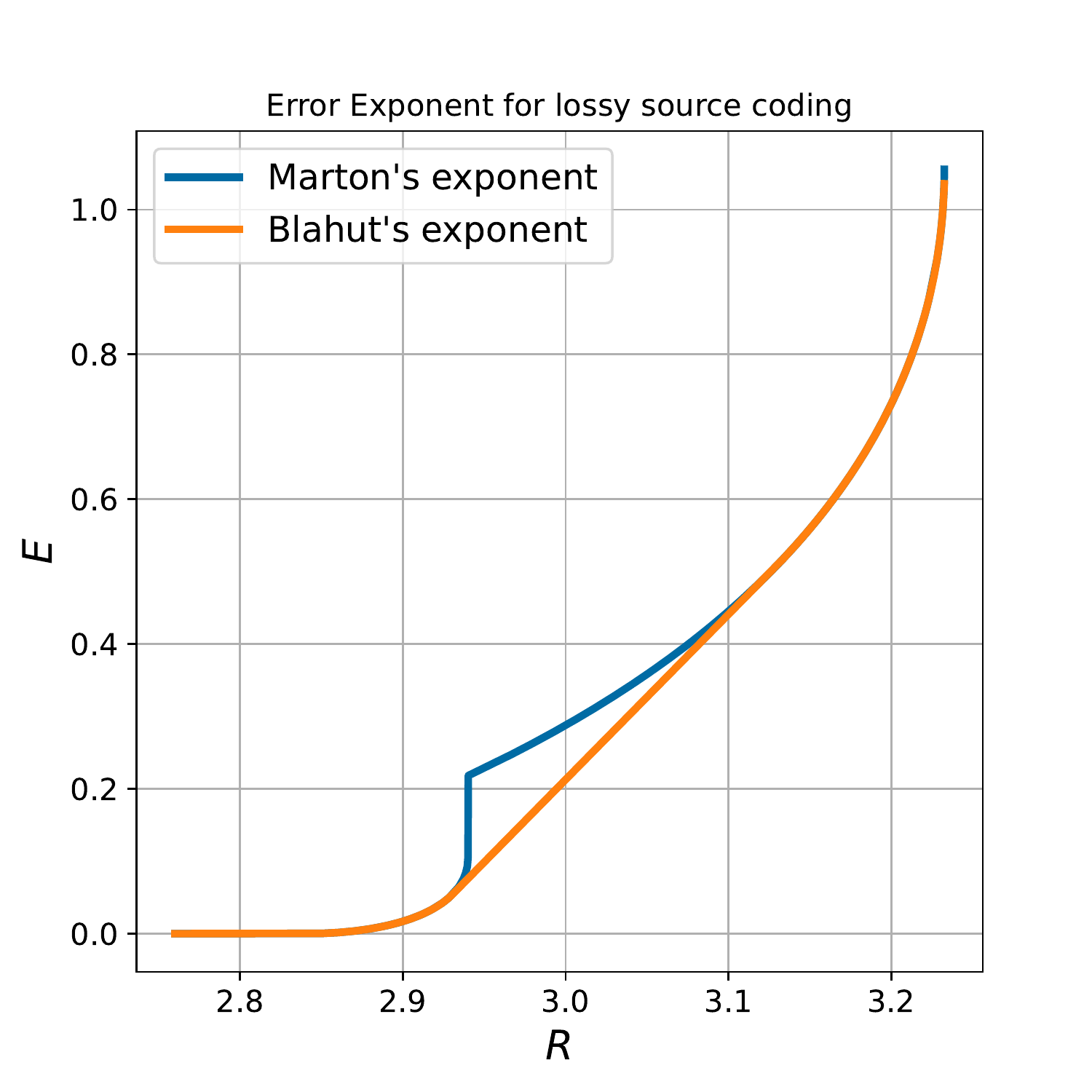}
    \caption{ Error exponents for the second example}
    \label{fig:error_exponent2}
\end{figure}

\begin{figure}
    \centering
    \includegraphics[width=0.8\columnwidth]{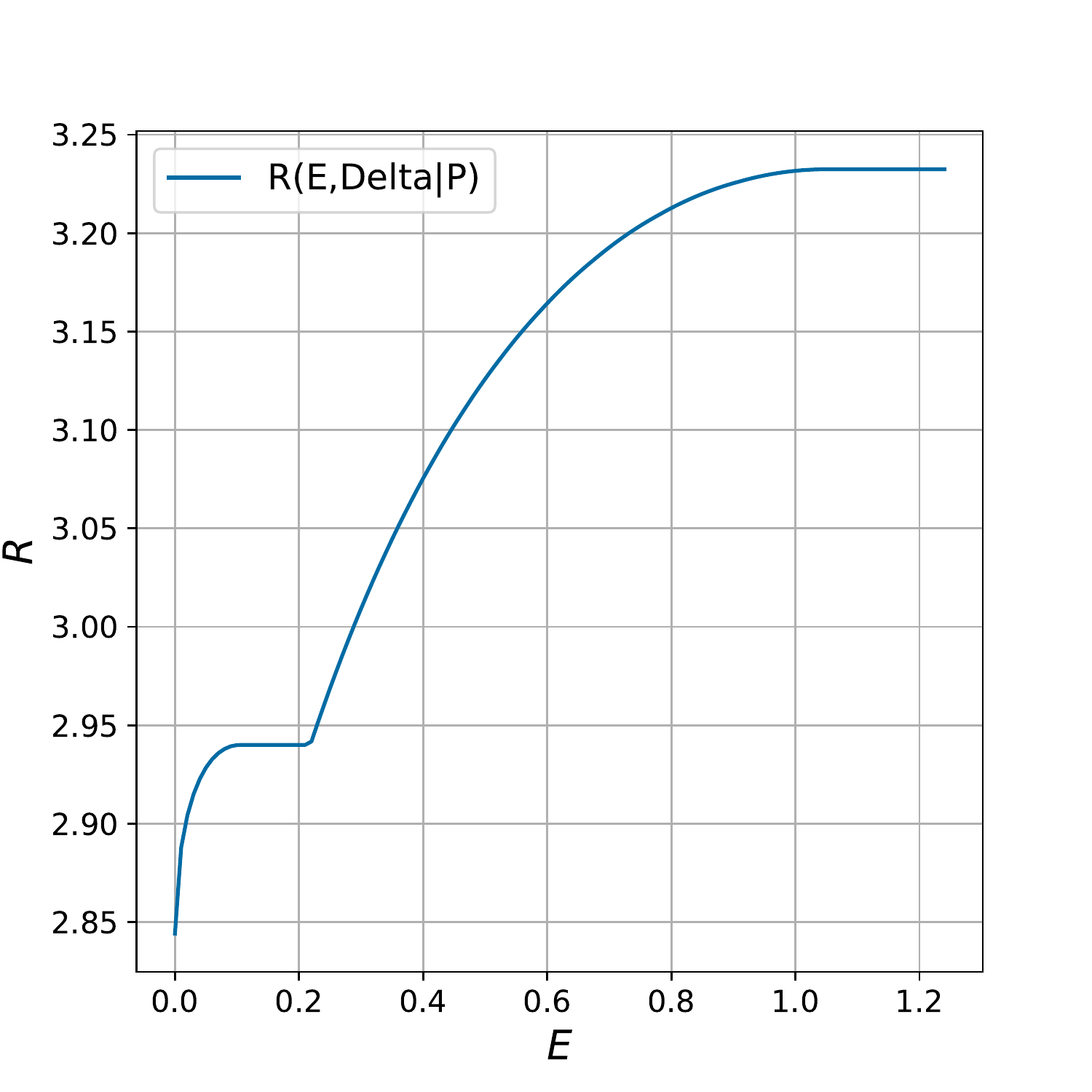}
    \caption{$R_{\rm M}(E | \Delta, P)$ for the second example }
    \label{fig:inverse2}
\end{figure}

\fi

\ifISIT
\else 
\section{Proofs of Lemmas~\ref{Lemma_G_nu} and \ref{lemma3}}
\label{Section:Proof}
In this section, we give the proofs of Lemmas~\ref{Lemma_G_nu} and \ref{lemma3} in Section III.

\textit{Proof of Lemma~\ref{Lemma_G_nu}:} 
If $\mu=0$, we have
\begin{align}
G^{(0,\nu)}(p_Y|P) 
&= \max_{q_X} 
-\sum_x q_X(x) \log \sum_y p_Y(y) {\rm e}^{-\nu d(x,y)} \notag\\
&= - \log \min_x \sum_y p_Y(y) {\rm e}^{-\nu d(x,y)}.
\end{align}
The maximum is attained by $q_X(x) = 1$ for
$x = \arg\min_x \sum_y p_Y(y) {\rm e}^{-\nu d(x,y)}$.
If $\mu>0$, we have
\begin{align}
& G^{(\mu,\nu)}(p_Y|P) \notag\\ 
&=
-\mu\min_{q_X} 
\left[
\sum_x q_X(x) \log \frac{q_X(x)}{ P(x) \left[ \sum_y p_Y(y) \mathrm{e}^{-\nu d(x,y) } \right]^{-1/\mu} }
\right] \notag \\
&= - \mu \min_{q_X} D(q_X||q_X^*) + \mu \log K \notag\\
&= \mu \log K, \notag
\end{align}
where 
$q_X^*(x) = \frac{1}{K} P(x) \left[ \sum_y p_Y(y) \mathrm{e}^{-\nu d(x,y) } \right]^{-1/\mu}$
and 
$K= \sum_x P(x) \left\{ \sum_{y} q_Y (y) \mathrm{e}^{-\nu d(x,y)} \right\}^{-1/\mu}$.
This completes the proof.\hfill\IEEEQED

Before describing the proof of Lemma~\ref{lemma3}, 
we show that the function $G^{(\nu)}(E, p_Y|P)$ satisfies the following property:
\begin{property} \label{property3}
For fixed $\nu\geq 0$, $p_Y$, and $P$, $G^{(\nu)}(E, p_Y | P)$ is a monotone non-decreasing and 
concave function of $E\geq 0$.
\end{property}

\

{\it Proof of Property~\ref{property3}}:
Monotonicity is obvious from the definition. Let us prove the concavity. 
Choose $E_0, E_1\geq 0$ arbitrarily. Set $E_\alpha = \alpha E_1 + (1-\alpha) E_0$
for $\alpha \in [0,1]$. Let the optimal distribution that attains 
$G^{(\nu)}(p_Y,  E_0 | P)$ and $G^{(\nu)}(p_Y,  E_1 | P)$
be $q_X^0$ and $q_X^1$. 
Then we have $D(q_X^i||P)\leq E_i$ for $i=0,1$. 
By the convexity of the KL divergence,
we have
$ D(\alpha q_X^1 + (1-\alpha) q_X^0 || P )
\leq
\alpha D(q_X^1  || P ) +
(1-\alpha) D(q_X^0 || P ) \leq
\alpha E_0 + (1-\alpha) E_1 = E_{\alpha}.
$
Therefore we have
\begin{align}
    & G^{(\mu)} (p_Y| E_\alpha, P) \notag\\
    & = \sup_{
    \genfrac{}{}{0pt}{}{q_X \in \mathcal{P(X)}:}{ D(q_X||P) \leq E_\alpha } 
    }
    \Big\{
     -\sum_x q_X(x) \log \sum_y p_Y(y) {\rm e}^{-\nu d(x,y) }
    \Big\} \notag\\
    &\geq 
    \left. -\sum_x q_X(x) \log \sum_y p_Y(y) {\rm e}^{-\nu d(x,y) } \right|_{q_X=  \alpha q_X^1 + (1-\alpha) q_X^0}\notag\\
&= \alpha G^{(\nu)}(p_Y, E_1|P) + (1-\alpha) G^{(\nu)}(p_Y, E_0|P) .
\end{align}
This completes the proof. \hfill$\IEEEQED$


\

\textit{Proof of Lemma~\ref{lemma3}}: 
For any $\mu\geq 0$, we have
\begin{align}
    & \sup_{
    \genfrac{}{}{0pt}{}{q_X \in \mathcal{P(X)}:}{ D(q_X||P) \leq E } 
    }
    \Big\{
     -\sum_x q_X(x) \log \sum_y p_Y(y) {\rm e}^{-\nu d(x,y) }
    \Big\} \notag\\
    & \leq    
    \sup_{
    \genfrac{}{}{0pt}{}{q_X \in \mathcal{P(X)}:}{ D(q_X||P) \leq E } 
    }
    \Big\{
     -\sum_x q_X(x) \log \sum_y p_Y(y) {\rm e}^{-\nu d(x,y) } \notag\\
    & \quad \quad+ \mu ( E-D(q_X||P) )
    \Big\} 
    \notag\\
    & \leq    
    \sup_{
        q_X \in \mathcal{P(X)}:
    }
    \Big\{
     -\sum_x q_X(x) \log \sum_y p_Y(y) {\rm e}^{-\nu d(x,y) } \notag\\
    &\quad \quad + \mu (E-D(q_X||P))
    \Big\}.
\end{align}
Thus, we have
    \begin{align}
        & G^{(\nu)} ( p_Y, E | P ) \notag\\
        & \leq 
        \inf_{\mu \ge 0}
        \bigg[
        \mu E + \max_{q_X\in \mathcal{P(X)}} 
        \Big\{
        - \mu D(q_X || P) \notag\\
        & \hspace{12mm} -\sum_x q_X(x) \log \sum_y p_Y(y) e^{ -\nu d(x,y) }
        \Big\}
        \bigg].
\end{align} 
Next, we prove that there exist a $\mu\geq0$ such that 
\begin{align}
&G^{(\nu)} ( p_Y, E | P ) \notag\\
&\geq\mu E + \max_{q_X\in \mathcal{P(X)}} 
        \{
        - \mu D(q_X || P)  \notag\\
        &\quad -\sum_x q_X(x) \log \sum_y p_Y(y) e^{ -\nu d(x,y) }
        \}. \label{eq.29}
\end{align}
From Property~\ref{property3}, for a fixed $E\geq 0$, there exist a $\mu \geq 0$
such that for any $E'$ we have
\begin{align}
    G^{(\nu)}(p_Y, E'|P) \leq G^{(\nu)}(p_Y, E|P) + \mu (E'-E). \label{eq.36}
\end{align}
Fix this $\mu=\mu(E)$ and put $E'=D(q_X'||P)$ for some $q_X'$.
Then, we have
\begin{align}
    & G^{(\nu)}(p_Y,  E' | P) \notag\\
    & = 
\sup_{
    \genfrac{}{}{0pt}{}{q_X \in \mathcal{P(X)}: }{ D(q_X||P) \leq E' } 
    }    \Big\{
     -\sum_x q_X(x) \log \sum_y p_Y(y) {\rm e}^{-\nu d(x,y) }
    \Big\}\notag\\
    &\geq 
     -\sum_x q_X'(x) \log \sum_y p_Y(y) {\rm e}^{-\nu d(x,y) } \label{eq.37}
\end{align}
Then, we have
\begin{align}
    & G^{(\nu)}(p_Y, E|P)  \notag\\
    &\
    \stackrel{\rm (a)}
    \geq G^{(\nu)}(p_Y, E'|P)+ \mu (E-E') \notag\\
    &
    \stackrel{\rm (b)}
    \geq      -\sum_x q_X'(x) \log \sum_y p_Y(y) {\rm e}^{-\nu d(x,y) } \notag\\
&\quad  + \mu E - \mu D(q_X'||P).
\end{align}
Step (a) follows form (\ref{eq.36}) and step (b) follows from 
(\ref{eq.37}) and the choice of $E'$. Thus for this choice of $\mu$, (\ref{eq.29}) holds.
This completes the proof. \hfill$\IEEEQED$

\fi

\section*{Acknowledgments}
The author thanks to Professor Yasutada Oohama
and Dr. Yuta Sakai for valuable comments. 
He also thanks to the anonymous reviewers for the helpful comments.  
A part of this work was supported by JSPS KAKENHI Grant Number JP19K12156 and JP23H01409. 

\bibliographystyle{IEEEtran}
\bibliography{mybibliography.bib, gyoseki.bib}

\ifISIT
\else 
\appendices
\section{Graph for Remark~\ref{remark3}}
\label{appendix_graph}
In Remark~\ref{remark3}, it was stated that
$ \max_{p_Y} E_{0,s}^{(\rho, \nu)}(p_X|P) $ is not necessarily concave 
in $\nu$. Here, we give an example to demonstrate
that nonlinear optimization over $\nu$ is required to
evaluate the Blahut's exponent. 
Ahlswede's counterexample with $|\mathcal{X}_A|=8$ and $|\mathcal{X}_B|=512$ is used and we put $\rho = 2.25$.
The graph in Fig.~\ref{fig.nonconvexity} shows $\max_{p_Y} E_{0,s}^{(\rho,\nu)} (p_Y|Q_\lambda) - \rho \nu \Delta$ against $\nu$, where optimal $p_Y$ is computed by Algorithm~\ref{algorithm1}.
This figure clearly shows that there are two local maxima.

\begin{figure}
    \centering
    \includegraphics[width=0.8\columnwidth]{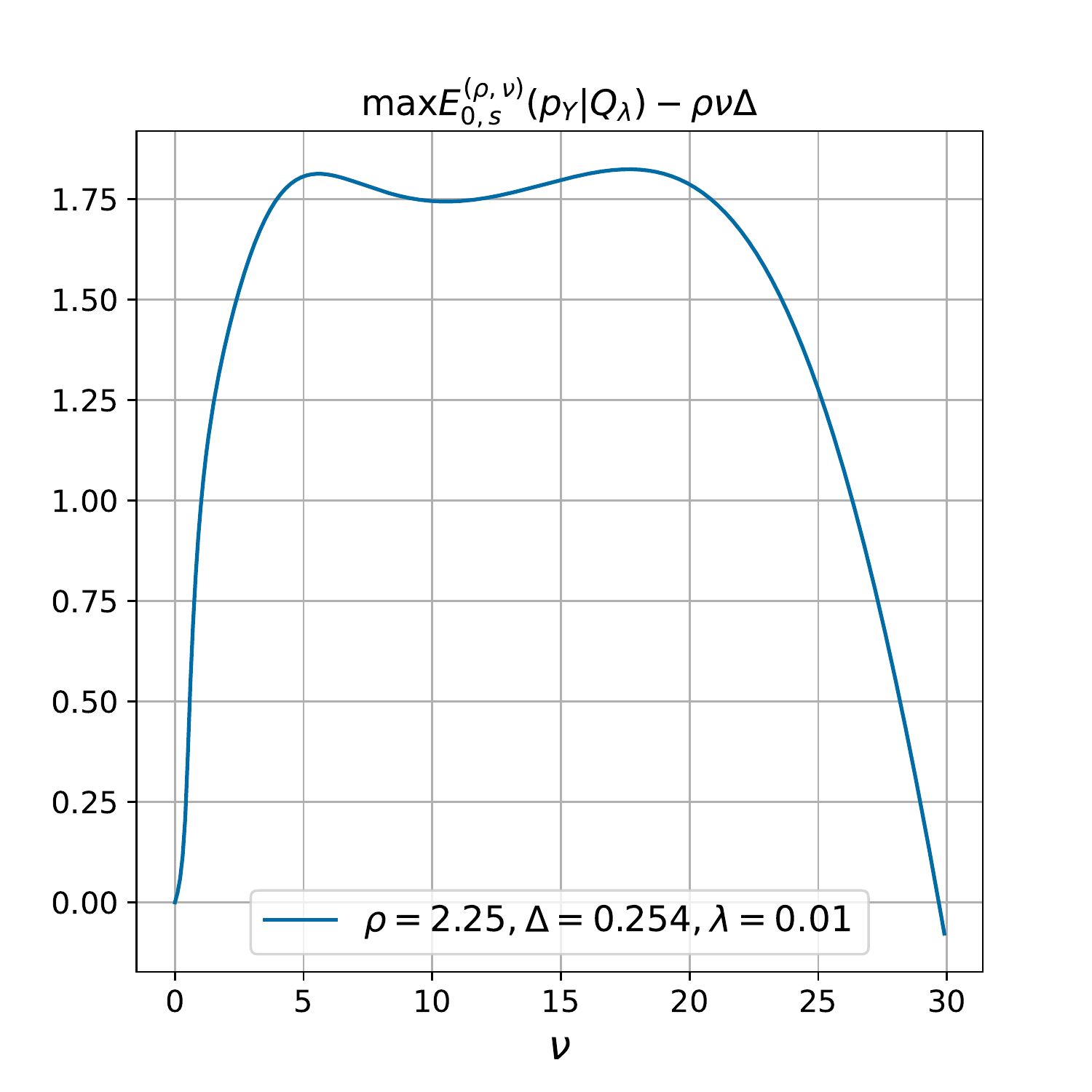}
    \caption{$\max_{p_Y} E_{0,s}^{(\rho,\nu)} (p_Y|Q_\lambda) - \rho \nu \Delta$ as a function of $\nu$}
    \label{fig.nonconvexity}
\end{figure}

\section{Proofs of lemmas~\ref{lemma.1} and~\ref{lemma4}}
\label{appendix_proof}
In this appendix, we give the proofs for the lemmas.

{\it Proof of Lemma~\ref{lemma.1}}:
Let $q_X^*$ be optimal distribution that achieves
$ E_{\rm M}(R|\Delta, P)  = \min_{q_X: R(\Delta|q_X)\geq R} D(q_X||P) $
and $\rho$ be any non-negative number. 
Then, we have 
\begin{align}
    & E_{\rm M}(R|\Delta, P) = D(q_X^*||P) \notag\\
    & \stackrel{\rm (a)}
      \ge \{ D(q_X^*||P) -\rho[R(\Delta|q_X^*)- R] \} \notag \\
    & \ge \min_{q_X: R(\Delta|q_X)\geq R} \{ D(q_X||P) -\rho[R(\Delta|q_X)- R] \}\notag\\
    & \ge \min_{q_X \in \mathcal{P(Y)} } \{ D(q_X||P) -\rho[R(\Delta|q_X)- R] \}\notag\\
    & \stackrel{\rm (b)}
      = \rho R + \min_{q_X\in \mathcal{P(Y)}} \Big\{ D(q_X||P) -\rho \sup_{\nu\ge 0} 
      \Big[ -\nu \Delta \notag\\
    & \quad - \max_{p_Y} \sum_x q_X(x) \log p_Y(y) {\rm e}^{-\nu d(x,y)}
    \Big] \Big\} \notag\\
    & = \rho R + \inf_{\nu\ge 0} \min_{q_X} \max_{p_Y} \Big[ 
    \rho\nu \Delta + D(q_X||P)    
      \notag\\
    & \hspace{2cm} + \rho\sum_x q_X(x) \log \sum_{y} p_Y(y) {\rm e}^{-\nu d(x,y)}  
    \Big] \notag\\
    & \stackrel{\rm (c)}
      = \rho R + \inf_{\nu\ge 0} \max_{p_Y} \min_{q_X} \Big[ 
      \rho\nu \Delta + D(q_X||P)    
      \notag\\
    & \hspace{2cm} + \rho\sum_x q_X(x) \log \sum_{y} p_Y(y) {\rm e}^{-\nu d(x,y)}  
    \Big] \notag\\
    & \stackrel{\rm (d)}
      = \rho R + \inf_{\nu\ge 0} \bigg[ \rho\nu \Delta      
      \notag\\
    & \quad + \max_{p_Y}-\log \sum_{x} P(x) \Big\{\sum_y p_Y(y) {\rm e}^{-\nu d(x,y)} \Big\}^{-\rho}  
    \bigg]. \label{eq.proof.lemma1} 
\end{align}
Step (a) holds because $q^*_X$ satisfies $R(\Delta|q_X) \ge R$. 
In Step (b), Eq.(\ref{parametric_Rate_distortion}) is substituted.
Step (c) follows from the minimax theorem. It holds 
because $D(q_X||P)$ is a convex function of $q_X$ and
$ \sum_x q_X(x) \log p_Y(y) {\rm e}^{-\nu d(x,y)}$ is linear in $q_X$ and
concave in $p_Y$. 
Step (d) holds because we have
\begin{align}
&    D(q_X||P) + \rho\sum_x q_X(x) \log \sum_{y} p_Y(y) {\rm e}^{-\nu d(x,y)} \notag\\
&= \sum_x q_X(x) \log \frac{q_X(x)}{ P(x) \{ \sum_{y} p_Y(y) {\rm e}^{-\nu d(x,y)} \}^{-\rho} } \notag\\
&=\sum_x q_X(x) \log \frac{q_X(x)}{ \frac{1}{K} P(x) \{ \sum_{y} p_Y(y) {\rm e}^{-\nu d(x,y)} \}^{-\rho} }
-\log K\notag \\
&\stackrel{\rm (e)}
\ge -\log K,
\end{align}
where $K = \sum_{x\in \mathcal{X} } P(x) \{ \sum_{y} p_Y(y) {\rm e}^{-\nu d(x,y)} \}^{-\rho} $.
In Step (e), equality holds when $q_X(x) = \frac{1}{K} P(x) \{ \sum_{y} p_Y(y) \cdot {\rm e}^{-\nu d(x,y)} \}^{-\rho}$. 
Because Eq.~(\ref{eq.proof.lemma1}) holds any $\rho\ge 0$, we have
\begin{align}
    & E_{\rm M}(R|\Delta, P) \notag\\
    & \ge \sup_{\rho\ge 0} 
    \bigg\{
\rho R + \inf_{\nu\ge 0} \bigg[ \rho\nu \Delta      
      \notag\\
    & \quad + \max_{p_Y} - \log \sum_{x} P(x) \Big\{\sum_y p_Y(y) {\rm e}^{-\nu d(x,y)} \Big\}^{-\rho}  
    \bigg]
    \bigg\}\notag\\
 & = E_{\rm B}(R|\Delta, P_X).    
\end{align}
This completes the proof.\hfill$\IEEEQED$

\textit{Proof of Lemma~\ref{lemma4}}:
The expression (\ref{parametric_Rate_distortion}) of the rate distortion function 
is related to the double minimization form of the Arimoto-Blahut algorithm.
We have the following chain of equations.
\begin{align}
& R(\Delta | q_X) \notag \\
& =
\min_{
\genfrac{}{}{0pt}{}{q_{Y|X} \in \mathcal{P(Y|X)}: }{
\mathrm{E}[d(X,Y)] \leq \Delta
}
}
I(q_X,q_{Y|X}) \notag\\
&=
\sup_{\nu \geq 0}
\bigg[ 
\min_{ q_{Y|X} \in \mathcal{P(Y|X)}: }
\big\{
I(q_X,q_{Y|X}) 
+ \nu \mathrm{E}_{q_{XY}} [ d(X,Y) ]
\big\} \notag \\
& \hspace{12mm} -\nu \Delta
\bigg]
\notag\\
&=
\sup_{\nu \geq 0}
\bigg[ 
\min_{ q_{Y|X} \in \mathcal{P(Y|X)}: }
\mathrm{E}_{q_{XY}} \left[ 
\log 
\frac{ q_{Y|X} (Y|X) }{ q_Y(Y) \mathrm{e}^{-\nu d(X,Y)} }
\right]
\notag\\
&\hspace{12mm} 
+ \min_{p_Y\in \mathcal{P(Y)}}
D(q_Y || p_Y) -\nu \Delta
\bigg]
\notag\\
& 
=
\sup_{\nu \geq 0}
\bigg[ 
\min_{ q_{Y|X} \in \mathcal{P(Y|X)} }
\min_{ p_Y \in \mathcal{P(Y)} } 
\mathrm{E}_{q_{XY}} \left[ 
\log 
\frac{ q_{Y|X} (Y|X) }{ p_Y(Y) \mathrm{e}^{-\nu d(X,Y)} }
\right] \notag\\
&\hspace{12mm} -\nu \Delta
\bigg] \label{Eq.41}
\end{align}
The double minimization in (\ref{Eq.41}) w.r.t. $p_Y$ and $q_{Y|X}$ is used to derive the Arimoto-Blahut algorithm. 
Let $A(x) = \sum_y p_Y(y) {\rm e}^{-\nu d(x,y)}$ and
$ q^{*}_{Y|X} (y|x) 
= A^{-1}(x) p_Y(y) {\rm e}^{-\nu d(x,y)}
$. 
Then, for a fixed $p_Y\in \mathcal{P(Y)}$, we have
\begin{align}
&
\min_{ q_{Y|X} \in \mathcal{P(Y|X)}: }
\mathrm{E}_{q_{XY}} \left[ 
\log 
\frac{ q_{Y|X} (Y|X) }{ p_Y(Y) \mathrm{e}^{-\nu d(X,Y)} }
\right] \notag \\
&
=
\min_{ q_{Y|X} \in \mathcal{P(Y|X)}: } 
\bigg[
\mathrm{E}_{q_{XY}} \left[ 
\log 
\frac{ q_{Y|X} (Y|X) }{ q_{Y|X}^*(Y|X) }
\right] \notag \\
& \hspace{12mm} - \mathrm{E}_{q_{X}} \left[ \log
A(X) 
\right] \bigg]\notag\\
&
=
\min_{ q_{Y|X} \in \mathcal{P(Y|X)}: } 
D(q_{Y|X} || q_{Y|X}^* | q_X) 
- \mathrm{E}_{q_{X}} \left[ \log
A(X) 
\right] \notag\\
&
\stackrel{ \rm (a) }
= 
- \mathrm{E}_{q_{X}} \left[ \log
A(X) 
\right]  \notag\\
&=
-\sum_{x} q_X(x) \log \sum_y p_Y(y) {\rm e}^{-\nu d(x,y)}.
\end{align}
In Step (a), $D(q_{Y|X} || q_{Y|X}^* | q_X) $ takes zero if and only if $q_{Y|X} =q_{Y|X}^*$,
which leads to the probability updating rule for the Arimoto-Blahut algorithm.
Thus, we have
\begin{align}
&
\min_{ q_{Y|X} \in \mathcal{P(Y|X)}: }
\min_{ p_{Y} \in \mathcal{P(Y)}: }
\mathrm{E}_{q_{XY}} \left[ 
\log 
\frac{ q_{Y|X} (Y|X) }{ p_Y(Y) \mathrm{e}^{-\nu d(X,Y)} }
\right] \notag\\
&=
\min_{ p_Y \in \mathcal{P(Y)} } 
-\sum_{x} q_X(x) \log \sum_y p_Y(y) {\rm e}^{-\nu d(x,y)}.
\label{eq.47}
\end{align}
Substituting (\ref{eq.47}) into (\ref{Eq.41}) yields 
\begin{align}
R(\Delta|q_X)
&=
\sup_{\nu \ge 0} \bigg[
\min_{p_Y \in \mathcal{P(Y)}}
-\sum_x q_X(x) \log \sum_y p_Y(y) {\rm e}^{-\nu d(x,y)} \notag\\
& \hspace{12mm} - \nu \Delta
\bigg].
\end{align}
This completes the proof.
\hfill\IEEEQED

\fi

\end{document}